\documentclass[12pt]{iopart}
\usepackage{iopams,epsfig,wrapfig} 

\newcommand{\eqeqref}[1]{Eq.~(\ref{#1})}

\newcommand{\refref}[1]{Ref.~\cite{#1}}

\newcommand{\beq}{\begin{equation}}
\newcommand{\eeq}{\end{equation}}
\newcommand{\bea}{\begin{eqnarray}}
\newcommand{\beas}{\begin{eqnarray*}}
\newcommand{\beau}[1]{\begin{equation} \label{#1} \begin{array}{rcl}}
\newcommand{\eea}{\end{eqnarray}}
\newcommand{\eeas}{\end{eqnarray*}}
\newcommand{\eeau}{\end{array} \end{equation}}
\newcommand{\bay}{\begin{array}}
\newcommand{\eay}{\end{array}}
\newcommand{\bals}{\begin{align*}}
\newcommand{\eals}{\end{align*}}

\newcommand{\ds}{\displaystyle}

\newcommand{\vev}[1]{\langle #1 \rangle}

\newcommand{\In}[1]{ \bigg| _{#1} }

\begin{document}


\title{Cronin effect and geometrical shadowing in d+Au collisions: pQCD vs. CGC}

\author{Alberto Accardi and Miklos Gyulassy}

\address{
Columbia University, Department of Physics\\
538 West 120th Street, New York, NY 10027, USA
}

\begin{abstract}
Multiple initial state parton interactions in p(d)+A collisions
are calculated in a Glauber-Eikonal formalism, which incorporates
the competing pattern of low-$p_T$ suppression due geometrical shadowing,
and a moderate-$p_T$ Cronin enhancement of hadron spectra. 
Dynamical shadowing effects, which are not included in the
computation, may be extracted by comparing experimental data to the
baseline provided by the Glauber-Eikonal model. 
Data for $\pi^0$ production at midrapidity show absence of dynamical
shadowing in the RHIC energy range, $\sqrt{s}\sim 20-200$ GeV. 
Recent preliminary data at forward rapidity are addressed,
and their interpretation discussed.
\end{abstract}





In proton ($p$), or deuteron ($d$), reactions involving heavy nuclei
($A$$\sim$$200$) at  $\sqrt{s}<40$ AGeV, the moderate transverse momentum
($p_T$$\sim$2-6 GeV) spectra are enhanced relative to linear extrapolation from 
$p+p$ reactions. This ``Cronin effect''
\cite{Cronin} is generally attributed to multiple
scattering of projectile partons propagating through the target
nucleus \cite{Accardi02}.
In this talk, we discuss multiple parton scattering in the
Glauber-Eikonal (GE) approach \cite{AG03b,GEmodels},
in which sequential multiple partonic collisions are computed in pQCD,
and unitarity is naturally
preserved. The low-$p_T$ spectra in $p+A$ collisions 
are suppressed by unitarity. At moderate $p_T$, the accumulation  of
transverse momentum  leads to an
enhancement of transverse spectra. At high $p_T$ the binary scaled
$p+p$ spectrum is recovered: no high-$p_T$ shadowing is
predicted in this approach.  \\

{\bf Hadron production in p+p collisions.} 
The first step to understand $p+A$ collisions is to understand $p+p$
collisions. The pQCD formula for the single inclusive hadron transverse
spectrum is:
\begin{equation}
  \fl \frac{d\sigma}{dp_T^2 dy}^{\hspace{-0.3cm}pp'\rightarrow hX}
    \hspace{-0.7cm} = \sum_{i=q,g}\Bigg\{
    \vev{xf_{i/p}}_{y_i,p_T} \, \frac{d\sigma^{\,ip'}}{dy_i d^2p_T} \In{y_i=y}
  + \vev{xf_{i/p'}}_{y_i,p_T} \, 
    \frac{d\sigma^{\,ip}}{dy_i d^2p_T}\In{y_i=-y} \Bigg\}
    \otimes D_{i\rightarrow h}(z,Q_h^2) \ .
 \label{ppcoll_pt}
\end{equation}
Here we considered only elastic parton-parton subprocesses,
which contribute to more than 98\% of the cross section at midrapidity.
In \eqeqref{ppcoll_pt}, 
\begin{eqnarray}
  \fl \vev{xf_{i/p}}_{y_i,p_T} &=& \ds {K \over \pi} 
      \sum_j \frac{1}{1+\delta_{ij}}  
      \int dy_2 x_1f_{i/p}(x_1,Q_p^2) 
      \frac{d\hat\sigma}{d\hat t}^{ij} \hspace{-0.2cm} 
      (\hat s,\hat t,\hat u) \, x_2f_{j/p'}(x_2,Q_p^2) 
    \Bigg/ 
      \frac{d\sigma^{ip'}}{d^2p_T dy_i}
 \label{avflux}
    \\
  \fl \frac{d\sigma^{ip'}}{d^2p_T dy_i} &=& {K \over \pi} 
    \sum_j \frac{1}{1+\delta_{ij}} 
    \int dy_2 \frac{d\hat\sigma}{d\hat t}^{ij} \hspace{-0.2cm} 
    (\hat s,\hat t,\hat u) \, x_2f_{j/p'}(x_2,Q_p^2) 
 \label{iNxsec}
\end{eqnarray}
are interpreted, respectively, as the average flux of incoming partons
of flavour $i$ from the hadron $p$, and the cross section for the
parton-hadron scattering.
The rapidities of the $i$ and $j$ partons in the final state 
are labelled by $y_i$ and $y_2$. 
Infrared regularization is performed by adding a small mass to the
gluon propagator and defining the Mandelstam variables 
$\hat t (\hat u) = - m_T^2 (1+e^{\mp y_i \pm y_2})$, with 
$m_{T}=\sqrt{p_T^2+p_0^2}$. For more details, see \refref{AG03b}.
Finally, inclusive hadron production is computed as a convolution of 
Eq.~(\ref{ppcoll_pt}) with a fragmentation function 
$D_{i\rightarrow h}(z,Q_h^2)$.

In Eqs.~(\ref{ppcoll_pt})-(\ref{iNxsec}), we
have two free parameters, $p_0$ and the K-factor $K$, and a somewhat arbitrary
choice of the factorization and fragmentation scales, $Q_p=Q_h=m_T/2$.
After making this choice, we fit $p_0$ and $K$ 
to hadron production data in $pp$ collisions at the
energy and rapidtiy of interest. 
Equation (\ref{ppcoll_pt}) is very satisfactory for $q_T\gtrsim 5$ GeV,  
but overpredicts the curvature of the hadron spectrum in the
$q_T$=1-5 GeV range. This can be corrected for by considering an
intrinsic transverse momentum, $k_T$,  for the colliding partons
\cite{Owens87}. We found that a fixed $\vev{k_T^2}=0.52$ GeV 
leads to a dramatic improvement in the computation, 
which now agrees  with data at the $\pm$40\% level \cite{AG03b}. 
Finally, we obtain, at midrapidity $\eta=0$, 
$p_0=0.7\pm0.1$ GeV and $K=1.07\pm 0.02$ at Fermilab, and
$p_0=1.0\pm0.1$ GeV and $K=0.99\pm0.03$ at RHIC. \\


{\bf From p+p to p+A collisions.}
Having fixed all parameters in p+p collisions, and defined the 
parton-nucleon cross section (\ref{iNxsec}), the 
GE expression for a parton-nucleus scattering is \cite{AT01b}:
\begin{eqnarray}
  \fl \frac{d\sigma^{\,iA}}{d^2p_Tdyd^2b}
     = \sum_{n=1}^{\infty} \frac{1}{n!} \int & d^2b \, d^2k_1 \cdots d^2k_n
     \, \delta\big(\sum _{i=1,n} {\vec k}_i - {\vec p_T}\big) 
     \nonumber \\
  \fl & \times \frac{d\sigma^{\,iN}}{d^2k_1} T_A(b) 
     \times \dots \times  
     \frac{d\sigma^{\,iN}}{d^2k_n} T_A(b)
     \, e^{\, - \sigma^{\,iN}(p_0) T_A(b)} \,
       \ ,
 \label{dWdp}
\end{eqnarray}
where $T_A(b)$ is the target nucleus thickness function at impact
parameter $b$. The exponential factor in \eqeqref{dWdp} represents the
probability that the parton suffered no semihard scatterings after the
$n$-th one, and explicitly implements unitarity at the
nuclear level. Assuming that the partons from $A$ suffer only one
scattering on $p$ or $d$, we may generalize \eqeqref{ppcoll_pt} as follows, 
without introducing further parameters:
\begin{eqnarray}
  \fl \frac{d\sigma}{d^2p_T dy d^2b}^{\hspace{-0.5cm}pA\rightarrow iX}
    \hspace{-0.5cm}=\Bigg\{
    \vev{xf_{i/p}}_{y_i,p_T} \, \frac{d\sigma^{\,iA}}{d^2p_T dy_i d^2b} 
    \In{y_i=y}
  \hspace*{-.6cm} + \hspace{.1cm}
    T_A(b) \sum_{b}\, \vev{xf_{i/A}}_{y_i,p_T} \, 
    \frac{d\sigma^{\,ip}}{d^2p_T dy_i}  
    \In{y_i=-y} \Bigg\} \otimes D_{i\rightarrow h}
     \nonumber
\end{eqnarray}

Unitarity introduces
a suppression of parton yields compared to the binary scaled p+p
case. This is best seen in integrated parton yields:
$d\sigma^{iA}/{dyd^2b} \approx 1 - e^{\, - \sigma^{\,iN}(p_0)
T_A(b)}$. At low opacity  $\chi = \sigma^{\,iN}(p_0) T_A(b) \ll 1$, 
i.e.,when the number of 
scatterings per parton is small, the binary scaling is
recovered. However, at large opacity, $\chi \gtrsim 1$, the parton yield
is suppressed: $d\sigma^{iA}/{dyd^2b} \ll 1 < \sigma^{\,iN}(p_0) T_A(b)$. This
suppression is what we call ``geometrical
shadowing'', since it is driven purely by the geometry of
the collision through the thickness function $T_A$. As the integrated
yield is dominated by small momentum partons, geometrical shadowing is
dominant at low $p_T$. 

Beside the geometrical quark and gluon shadowing, which is automatically
included in GE models, at low enough $x$ one expects genuine dynamical
shadowing due to non-linear gluon interactions as described in 
Colour Glass Condensate (CGC) models \cite{CGC}. 
However, it is difficult to disentangle these two
sources of suppression of $p_T$ spectra, and to understand where
dynamical effects begin to play a role beside the ubiquitous
geometrical effects. The GE model computation outlined above can be
used as a baseline to  extract the magnitude of dynamical effects by
comparison with experimental data. \\

\begin{figure}[t]
\begin{center}
\epsfig{figure=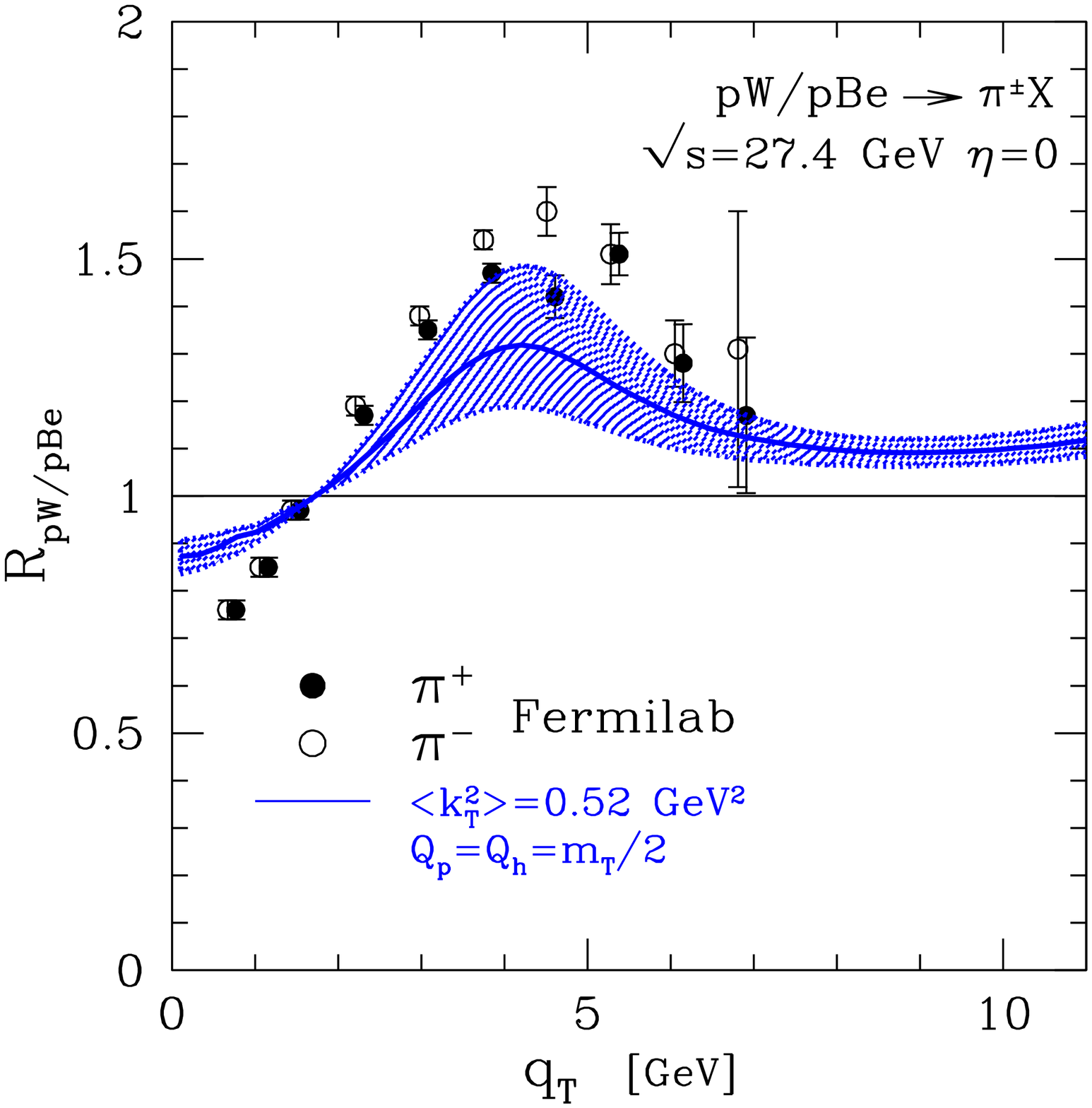,width=5cm}
\epsfig{figure=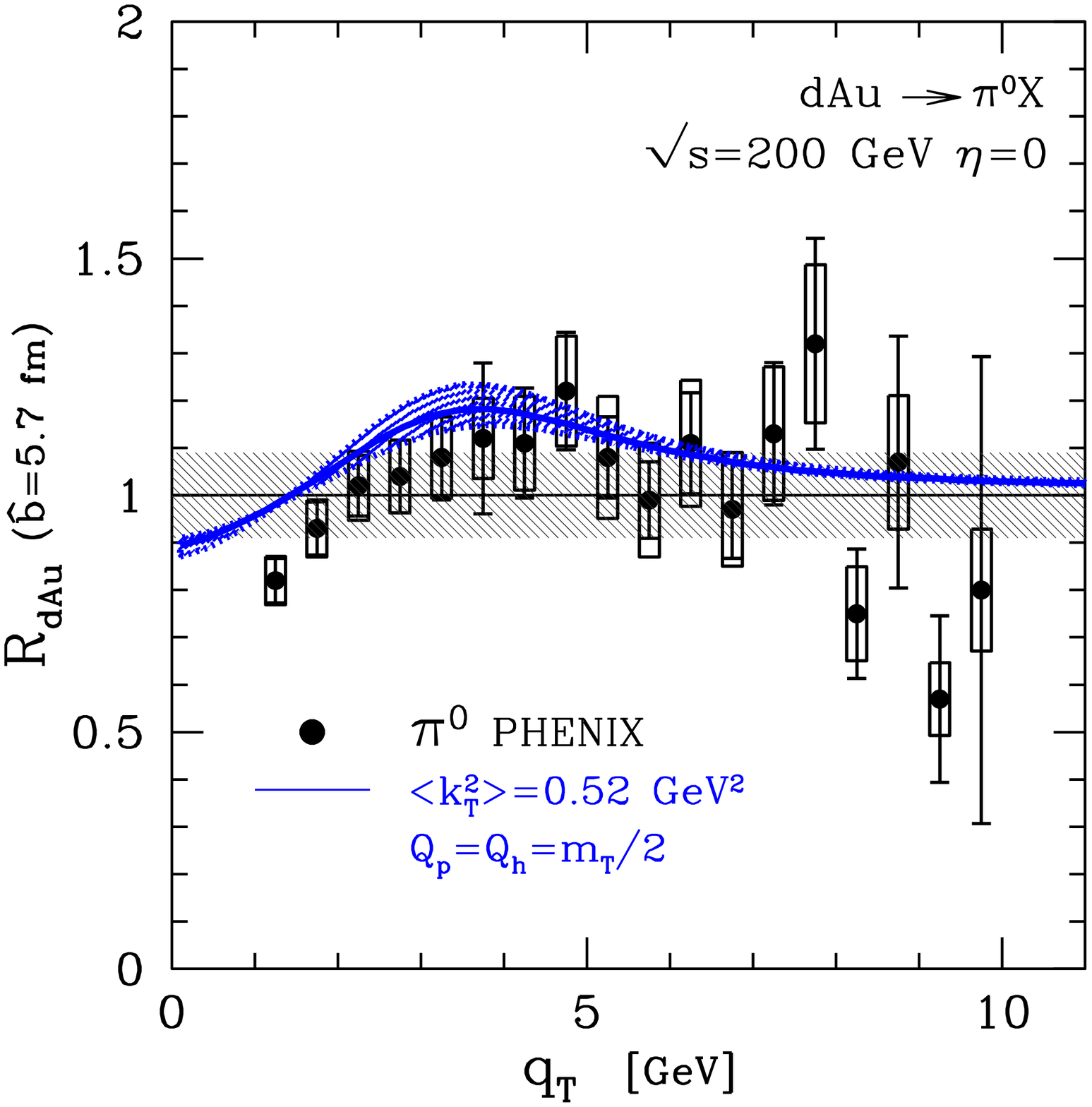,width=5cm}
\epsfig{figure=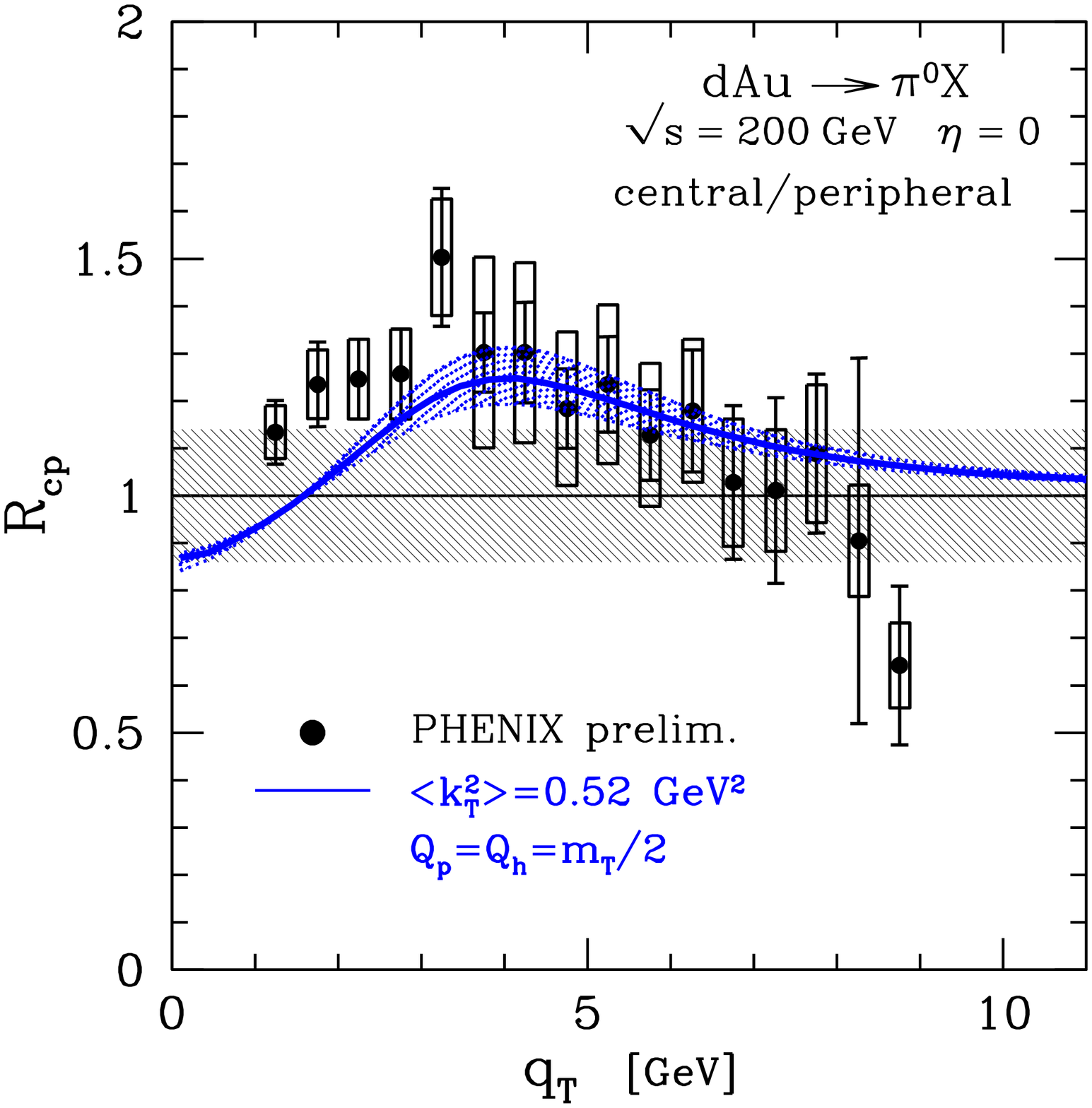,width=5cm}
\vskip-.2cm
\caption{\footnotesize Cronin effect on pion production at Fermilab
  \cite{Cronin2} and RHIC \cite{PHENIX} at $\eta=0$. The solid curve
  is the GE computation. Theoretical errors due to the fit of $p_0$
  are shown as a shaded band around the solid curve. The rightmost panel  
  shows the 0-20\%/60-88\% centrality classes ratio.} 
\label{fig:midrapidity}.
\end{center}
\vskip-1.2cm
\end{figure}

 
{\bf Cronin effect at Fermilab and RHIC.}
The Cronin effect may be quantified by taking the ratio of hadron
$p_T$ spectrum in p(d)+A collision, and dividing it by the binary
scaled p+p spectrum:
\begin{equation}
  R_{pA} \simeq 
    {\frac{d\sigma}{dq_T^2 dy d^2b}
       ^{\hspace{-0.5cm}dAu\rightarrow h X}
       \hspace{-0.8cm} (b=\hat b)} \Bigg/
    T_{A}(\hat b) {\frac{d\sigma}{dq_T^2 dy}
       ^{\hspace{-0.3cm}pp\rightarrow h X}
       \hspace{-0.8cm} } \ ,
  \label{CroninRatio}
\end{equation}
where $\hat b$ is the average impact parameter in the centrality bin
in which experimental data are collected (see \cite{AG03b}).
The GE model reproduces quite well both
Fermilab and PHENIX data at $\eta=0$ (Fig.1, left and center). 
It also describe the
increase of the Cronin effect with increasing centrality (Fig.1, right).
If dynamical shadowing as predicted in CGC models 
was operating in this rapidity and energy range, the
central/peripheral ratio should be smaller than the GE result: the
more central the collision, the higher the parton density in the
nucleus, the larger the non-linear effects. Therefore, we
conclude that there is no dynamical shadowing nor Colour Glass
Condensate at RHIC midrapidity.
   
To address the preliminary BRAHMS data at forward rapidity 
$\eta\approx3.2$ \cite{BRAHMSfwd}, we would first need to fit $p_0$
and $K$ in p+p collisions at the same pseudo-rapidity. 
Unfortunately the available data $p_T$-range
$0.5 \lesssim p_T \lesssim 3.5$ Ge is not large enough $p_T$ for the fit to
be done. Therefore, we use the parameters extracted at $\eta=0$. 
The resulting Cronin ratio, shown by the solid line in Figure 2,
overestimates the data at such low-$p_T$. 
This may be corrected in part by considering elastic energy loss 
\cite{CT04}. 
\begin{wrapfigure}{l}{6.3cm}
\label{fig:fwdrap}
\begin{center}
\vskip-.4cm
\parbox{6cm}{\epsfig{figure=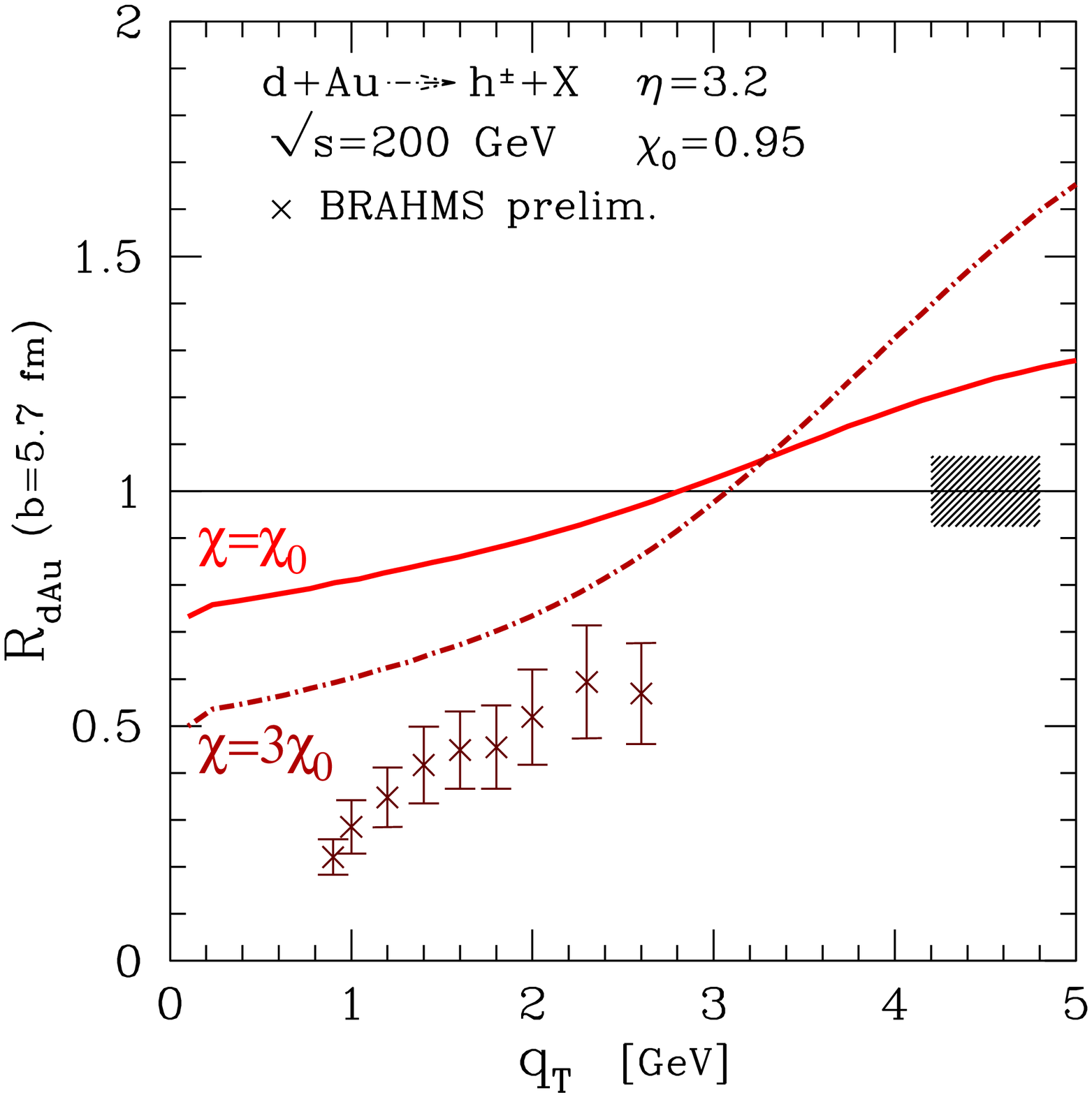,width=6cm}}\\[.1cm]
%
{\footnotesize {\bf Figure 2.}
Cronin ratio at $\eta=3.2$.}
\end{center}
\vskip-.4cm
\end{wrapfigure}
Furthermore, the opacity $\chi_0=0.95$ might be
underestimated, due to the use of the mid-rapidity parameters: to
check this we tripled the  opacity, but the resulting dashed line
still overestimates the data. 

Does the discrepancy between our calculations and the BRAHMS data prove that
a CGC has been observed? It is too early to tell. An important
observation \cite{BRAHMSfwd} is that the yield of 
positive charged hadrons exceeds the yield of negative charged hadrons 
at moderate pt by just the factor expected in HIJING
due to dominance of valence quark fragmentation effects: 
gluon fragmentation is subdominant mechanism in
this kinematic range. In addition,
current simplified CGC scenrarios predict asymptotic $1/p_T^4$ absolute
behavior which overestimates by an order of magnitude
the observed absolute cross sections. In our pQCD approach the absolute
cross sections in both p+p and d+Au collisions
are much closer to the data, though the modest discrepancies shown in Figure 2
remain. Gluon shadowing appears to be needed in this $x\sim
10^{-3}-10^{-2}$ regime, though more data will be required to quantify
the effect. 

The optimal region to study the onset of dynamical shadowing is $3
\lesssim p_T \lesssim 6$ GeV, where the GE model expects the 
peak of the Cronin ratio to be. If in this region the final
data at forward rapidity $0 \lesssim \eta \lesssim 4$ explored by the
four RHIC collaboration will reveal a consistent pattern of
suppression compared to the GE computation (as preliminary data seem
to suggest), and nonperturbative fragmentation effects will not be
able to explain it, then the case for the CGC will be made more solid.  
The case would be even stronger if at the same time 
a progressive disappearance of back-to-back jets in favour of an
increased dominance of monojet production was observed. 

\ \\[-.7cm]

\end{document}